\documentclass[pre,amsmath,amsfonts,amssymb,
aps,twocolumn,superscriptaddress,floatfix]{revtex4}
\usepackage{mathrsfs}
\usepackage{epsfig}
\usepackage{graphicx}

\usepackage[figuresright]{rotating}
\usepackage{bm}
\usepackage{color}
\usepackage{ulem}
\usepackage{environ}

\def\be{\begin{equation}} \def\ee{\end{equation}}
\def\bea{\begin{eqnarray}} \def\eea{\end{eqnarray}}

\newcommand{\WQC} {Wilczek Quantum Center and Key Laboratory of Artificial Structures and Quantum Control, School of Physics and Astronomy, Shanghai Jiao Tong University, Shanghai 200240, China}

\newcommand{\SRCQC}{Shanghai Research Center for Quantum Sciences, Shanghai 201315, China}

\begin{document}
\title{Emergent non-Hermitian physics in generalized Lotka-Volterra model}

\author{Tengzhou Zhang}
\affiliation{\WQC}


\author{Zi Cai}
\email{zcai@sjtu.edu.cn}
\affiliation{\WQC}
\affiliation{\SRCQC}

\begin{abstract}  
In this work, we study that non-Hermitian physics emerging from  a predator-prey ecological model described by a generalized Lotka-Volterra equation. In the phase space, this nonlinear equation exhibits both chaotic and localized dynamics, which are separated by a critical point. These distinct dynamics originate from the interplay between the periodicity and  non-Hermiticity of the effective Hamiltonian in the linearized equation of motion. Moreover, the dynamics at the critical point, such as algebraic divergence, can be understood as an exceptional point in the context of non-Hermitian physics.

\end{abstract}


\maketitle

\section{Introduction}

Physically, non-Hermitian Hamiltonians\cite{Ashida2020}, as a phenomenological description of process with energy or particle flowing out of the Hilbert space of interest, are responsible for  diverse intriguing phenomena in the contexts of classical and quantum waves\cite{Ruschhaupt2005,Ruter2010,Peng2014,Feng2014,Bertoldi2017,Xiao2020}, topological physics\cite{Lee2016,Yao2018,Yao2018b,Gong2018b,Liu2019,Borgnia2020,Bergholtz2021,Wang2020},  and active matters\cite{Fruchart2021}. Searching for physically transparent examples of non-Hermitian Hamiltonian is not only of fundamental interest for exploring non-Hermitian physics in a broader context, but also of practical significance due to its potential application in quantum sensing\cite{Chen2017,Hodaei2017} and  energy transfer\cite{Yu2017,Xu2016,Budich2020}.

In this study, we propose a generalized Lotka-Volterra equation (GLVE) in a one-dimensional (1D) lattice, which could exhibit chaotic or stable dynamics in different parameter regimes. The Lotka-Volterra (LV) equation  describing the predator-prey ecological processes  is a paradigmatic model in population dynamics\cite{Lotka1910,Volterra1928,Goel1971}. Recently, the GLVE has been generalized to spatially periodic systems to study the topological phases and edge modes beyond the scope of natural science\cite{Knebel2020,Yoshida2021,Tang2021}. The dynamics of a slight deviation from the stationary point of the GLVE are governed by a linearized equation resembling the single-particle Schrodinger equation in a lattice system. Therefore, the topological band theory can straightforwardly be applied to such a classical system\cite{Knebel2020,Yoshida2021,Umer2022}. Here, we show that if the linear expansion is performed around a temporal periodic solution instead of the stationary point of the GLVE, the equation of motion (EOM) of the deviation can also be described by the Schrodinger equation, but with a time-dependent non-Hermitian Hamiltonian. The exponential divergence to chaos and the stable, quasi-unitary dynamics both emerge from the Floquet quasi-energy band structure. The dynamical critical point in the original nonlinear model can be understood as an exceptional point of the non-Hermitian Floquet Hamiltonian.

\section{Model and method}

\subsection{The coupled predator-prey circles}

\begin{figure}[htb]
\includegraphics[width=0.99\linewidth]{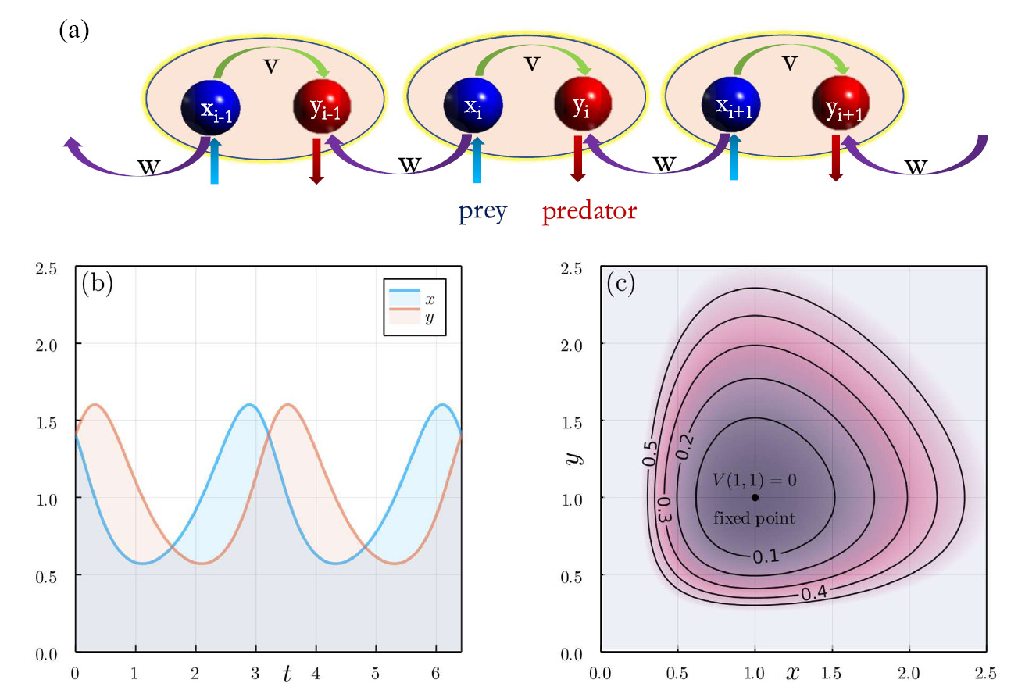}
\includegraphics[width=0.99\linewidth]{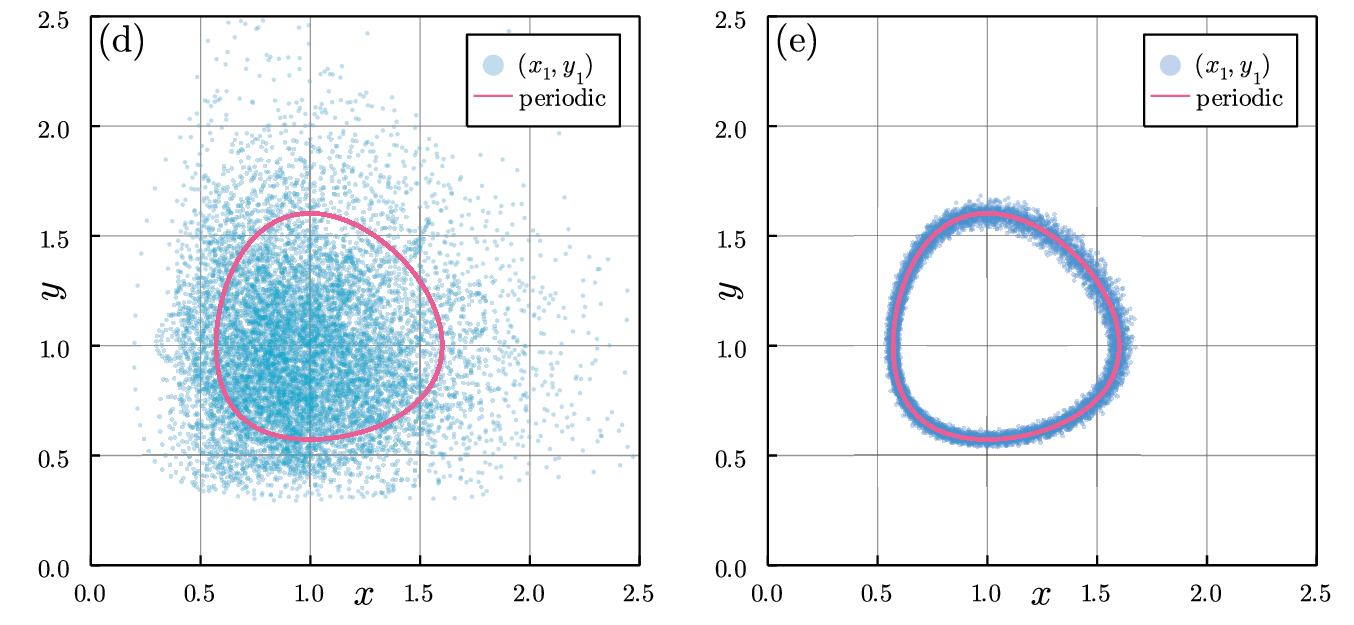}
\caption{(Color online)(a) Predator-prey model defined in a 1D ``diatomic''chain described by the GLVE Eq.(\ref{eq:LV}). (b)Periodic solution $[\bar{x}(t),\bar{y}(t)]^T$ of the homogeneous GLVE Eq.(\ref{eq:LV2}) with the conserved quantity $V=0.131$. (c) Trajectories of  $[\bar{x}(t),\bar{y}(t)]^T$ in the phase space with different conserved quantities. (d) and (e) Trajectories in the phase space of the first unit cell ($i=1$) predicted via the GLVE Eq.(\ref{eq:LV}) with (d) $r=0.3$ and (e) $r=0.7$, $\Delta=0.05$ and $L=1024$. The initial state of (d) and (e) is spatially inhomogeneous: $\delta_i(t=0)=\Delta_i$ with $\Delta_i$ being randomly sampled from $[-\Delta,\Delta]$.  The red curves indicate the trajectory starting from  the spatially homogeneous initial state $\delta_i(t=0)=0$.} \label{fig:Fig1}
\end{figure}

We focus on the GLVE defined in a 1D ``diatomic''chain (see Fig.\ref{fig:Fig1} a), which reads:
\begin{equation}\label{eq:LV}
\begin{split}
  \dot{x}_i&=x_i[2-v y_{i-1}-w y_i]\\
\dot{y}_i&=y_i[-2+v x_i+w x_{i+1}]
\end{split}\,,
\end{equation}
where $i=1\cdots L$, and $L$ is the number of the unit cell, each of which contains a prey ($x_i$) and predator ($y_i$). $v=1+r$ and $w=1-r$.  $0<r<1$ is the only tunable parameter in Eq.(\ref{eq:LV}) characterizing the difference between the inter and intra unit cell coupling strengths. The linear terms in the right side of Eq.(\ref{eq:LV}) suggest an exponential growth/decay for the prey/predator populations  if there is no interspecies interaction, while the nonlinear terms indicate the interaction between one specie and its neighbors, which suppress the exponential growth/decay.

Starting with a simple situation where the populations of prey and predator are site-independent $x_i(t)=x(t)$, $y_i(t)=y(t)$, Eq.(\ref{eq:LV}) is reduced to a two-species LV equation:
\begin{equation}\label{eq:LV2}
  \begin{split}
    \dot{x}&=2x-2 xy\\
    \dot{y}&=-2y+2xy
  \end{split}\, ,
\end{equation}
which is commonly used to explain the oscillation behavior of natural populations ({\it e.g.} the snowshoe hare and lynx) in ecological systems with predator-prey interactions, competition and disease. Mathematically, this model is integrable with a constant of motion\cite{Goel1971}, $V=x+y-\ln xy-2$. Consequently, it supports either a steady solution $[x^\star,y^\star]^T=[1,1]^T$ (with $V=0$) or a periodic oscillation $[\bar{x}(t),\bar{y}(t)]^T$ (with $V>0$) (see Fig.\ref{fig:Fig1} b), corresponding to a fixed point or a closed orbit around the fixed point in the phase space respectively (see Fig.\ref{fig:Fig1} c).

In general, one needs to take the spatial fluctuation into account. Considering a solution  $\pmb{v}(t)=[x_1,y_1,\cdots, x_L,y_L]^T$  of Eq.(\ref{eq:LV}), one can expand it around the spatially homogeneous solutions as
\begin{equation}\label{eq:expansion}
 v_i(t)=[1+\delta_i(t)]\bar{v}_i(t),
\end{equation}
where $\pmb\delta(t)=[\delta_1^x(t),\delta_1^y(t),\cdots]^T$  ($\delta_i^x(t)=\frac{x_i(t)-\bar{x}(t)}{\bar{x}(t)}$ and $\delta_i^y(t)$ is likewise). $\bar{\pmb{v}}$ donates a unperturbed solution and is not necessarily spatial homogeneous. A linearized equation can be derived in terms of the dimensionless vector $\pmb\delta(t)$.

\subsection{Linear expansion around the stationary solution}
For a homogeneous stationary solution $\bar{\pmb{v}}^\star(t)=[1,1,\cdots,1,1]^T$ , it is shown that  the linearized EOM of $\pmb\delta(t)$ takes the identical form of the single-particle Schrodinger equation in a 1D lattice:
\begin{equation}
i\frac{d \pmb\delta(t)}{dt}=H \pmb\delta(t), \label{eq:EOM0}
\end{equation}
where $H=H_0$ is a time-independent $2L\times 2L$ antisymmetric Hermitian matrix (due to the prefactor $i$):
\begin{equation}
H_0 =
    i\begin{bmatrix}
      0& -v  &   &   & -w    \\
      v& 0  & w  &   &     \\
       & -w & 0  & -v  &     \\
       &    &  v & 0  & \ddots    \\
      w&   &     & \ddots  & \ddots    \\
  \end{bmatrix}. \label{eq:H0}
\end{equation}

\subsection{Linear expansion around the periodic solution}

Unlike previous studies\cite{Knebel2020,Yoshida2021}, here we expand the nonlinear Eq.(\ref{eq:LV}) around the periodic solution $\bar{\mathbf{v}}_p(t)=[\bar{x}(t),\bar{y}(t),\cdots,\bar{x}(t),\bar{y}(t)]^T$, where $\bar{x}(t),\bar{y}(t)$ are the solution of Eq.(\ref{eq:LV2}) with a period $T\approx\pi$. The linearized EOM takes the same form of Eq.(\ref{eq:EOM0}),but with a time-dependent non-Hermitian ``Hamiltonian''
\begin{equation}
H(t)=H_0 D(t), \label{eq:nonHermitian}
\end{equation}
where $H_0$ has the same definition as Eq.(\ref{eq:H0}), and $D(t)$ is a diagonal matrix with dimension $2L$:
 \begin{equation}
D(t) =
    \begin{bmatrix}
      \bar{x}(t)&   &   &   &     \\
      & \bar{y}(t) &   &   &     \\
      &   &   \ddots &   &     \\
       &    &   &    \bar{x}(t) &     \\
      &   &     &  & \bar{y}(t)    \\
  \end{bmatrix}. \label{eq:H1}
\end{equation}

\begin{figure}[htb]
\includegraphics[width=1.0\linewidth]{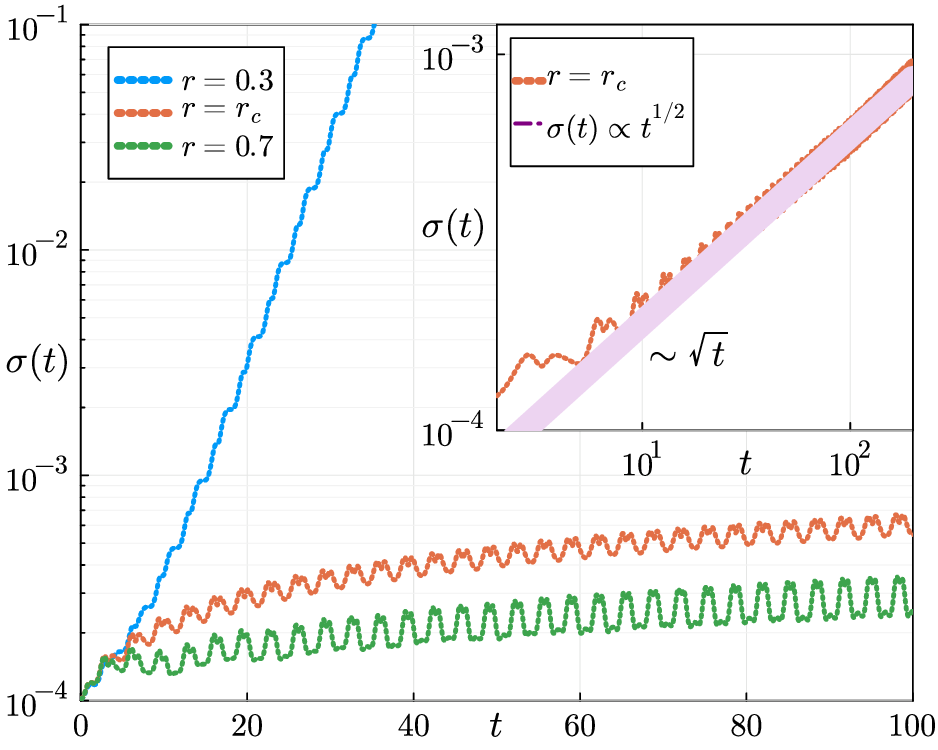}
\caption{(Color online) Dynamics of the average deviation $\sigma(t)$ with different $r$ values in a semi-log plot($r_c=0.64579$ is the critical point). The inset presents the dynamics of $\sigma(t)$ at the critical point in the log-log plot. The initial state is chosen as $x_i(t=0)=y_i(t=0)=1.6(1+\Delta_i)$ where the amplitude of the periodic solution $\xi \approx 0.33$ and $\Delta_i$ is randomly sampled from $[-\Delta,\Delta]$ where $\Delta=2\times 10^{-4}$. } \label{fig:Fig2}
\end{figure}

\section{Chaotic versus localized dynamics in the phase space}
Before discussing  the linearized EOM, we first focus on the dynamics of the nonlinear Eq.(\ref{eq:LV}), which can be solved using the standard Runge-Kutta method. A key question is whether the spatially homogeneous periodic solution $[\bar{x}(t),\bar{y}(t)]^T$ is stable against spatial fluctuations. To address this issue, we  impose a small site-dependent perturbation on the initial state as $\delta_i(t=0)=\Delta_i$, where $\Delta_i$ is randomly sampled from a uniform random distribution with $\Delta_i\in[-\Delta,\Delta]$ and $\Delta\ll 1$ (for a spatially homogeneous solution $\delta_i(t=0)=0$).  We first study the dynamics in one unit cell (say, $i=1$) by plotting the trajectories of $x_1(t)$ and $y_1(t)$ in the phase space. As shown in Fig.\ref{fig:Fig1} (d), for a small $r$ ({\it e.g.} $r=0.3$), the trajectory of  $[x_1(t),y_1(t)]^T$ rapidly deviates from the spatially homogeneous solution $[\bar{x}(t),\bar{y}(t)]^T$ after a short time, while randomly walking in the phase space on long timescales, indicating that the solution $[\bar{x}(t),\bar{y}(t)]^T$ is unstable against spatial fluctuation for small $r$. Conversely, at a relatively large $r$ ({\it e.g.} $r=0.7$), the trajectory of $[x_1(t),y_1(t)]$ is bounded within a finite regime around $[\bar{x}(t),\bar{y}(t)]$ (see  Fig.\ref{fig:Fig1} e).

The qualitatively different dynamical behavior between the cases with small and large values of $r$ reveal a non-equilibrium phase transition, which can be characterized by the average deviation: $\sigma(t)=\sqrt{\frac 1L \sum_i[\delta_i^x(t)]^2+[\delta_i^y(t)]^2}$. As shown in Fig.\ref{fig:Fig2}, $\sigma(t)$ increases exponentially (accompanied by an oscillation) at small $r$ (a signature of chaos), while it keeps oscillating around a finite value at a large $r$. The exponent of the exponential divergence approaches zero at  critical $r=r_c$, whose value depends on the amplitude of the periodic oscillation of the spatially homogeneous solutions. At the dynamical critical point,  $\sigma(t)$ grows algebraically as $\sigma(t)\sim t^{\frac 12}$. In the following, we will explain these observed dynamical behaviors as well as the critical dynamics based on the properties of the non-Hermitian Hamiltonian in Eq.(\ref{eq:nonHermitian}).

\section{Floquet dynamics with a non-Hermitian Hamiltonian}

Now we focus on the linearized EOM Eq.(\ref{eq:EOM0}) where the time-dependent Hamiltonian (\ref{eq:nonHermitian}) is non-Hermitian but  periodic in time $H(t)=H(t+T)$. However, unlike the intensively studied cases with periodically driven Hamiltonian, the periodic oscillation in Hamiltonian Eq.(\ref{eq:nonHermitian}) is not due to external driving, but originates from the spontaneous oscillation in the time-independent GLVE Eq.(\ref{eq:LV}), and is self-sustained. Thanks to the spatially translational invariance, one can perform the Fourier transformation, after which the EOM Eq.(\ref{eq:EOM0}) turns into a collection of independent $k$ modes, each of which is a two-level system governed by the EOM:
\begin{equation}
i\frac{d \pmb\delta_k}{dt}=H_k(t) \pmb{\delta}_k , \label{eq:EOMk}
\end{equation}
where $\pmb\delta_k=[\delta_k^x,\delta_k^y]^T$ with $
\delta_k^x=\frac 1{\sqrt{L}}\sum_j e^{-ik j}\delta^x_j$ and $\delta_k^y$ is likewise. $H_k$ is a $2\times2$ matrix defined as:
\begin{equation}
H_k(t) =H_k^0 D(t) ,\label{eq:Hk}
\end{equation}
with
\begin{small}
\begin{equation}
 H_k^0=\begin{bmatrix}
      0& -i(v+we^{-ik})      \\
      i(v+we^{ik})& 0
  \end{bmatrix},
  D(t)=\begin{bmatrix}
      \bar{x}(t)&       \\
      & \bar{y}(t)
  \end{bmatrix}. \label{eq:diag}
\end{equation}
\end{small}
Again, $H_k$ is non-Hermitian if  $\bar{x}(t)\neq  \bar{y}(t)$. Its instantaneous eigenvalues are still real but the dynamics is not trivial, since generally $[H_k(t_1),H_k(t_2)]\ne 0$. Both $\bar{x}(t)$ and  $\bar{y}(t)$ are periodic in time with a period $T$, enabling us to employ the Floquet description of the dynamics of Eq.(\ref{eq:EOMk}) and derive a time-independent Floquet Hamiltonian $H_k^F$ satisfying:
\begin{equation}
\mathcal{F}_k=e^{-iH_k^F T}=\mathcal{T}  e^{-i\int_0^T dt H_k(t)},
\label{eq:Floquet0}
\end{equation}
where $\mathcal{T}$ is the time-ordering operator and $\mathcal{F}_k$ is the evolution operator for the k-mode within one period whcih is not necessarily unitary\cite{Wu2020}.

\begin{figure}[htb]
\includegraphics[width=0.99\linewidth]{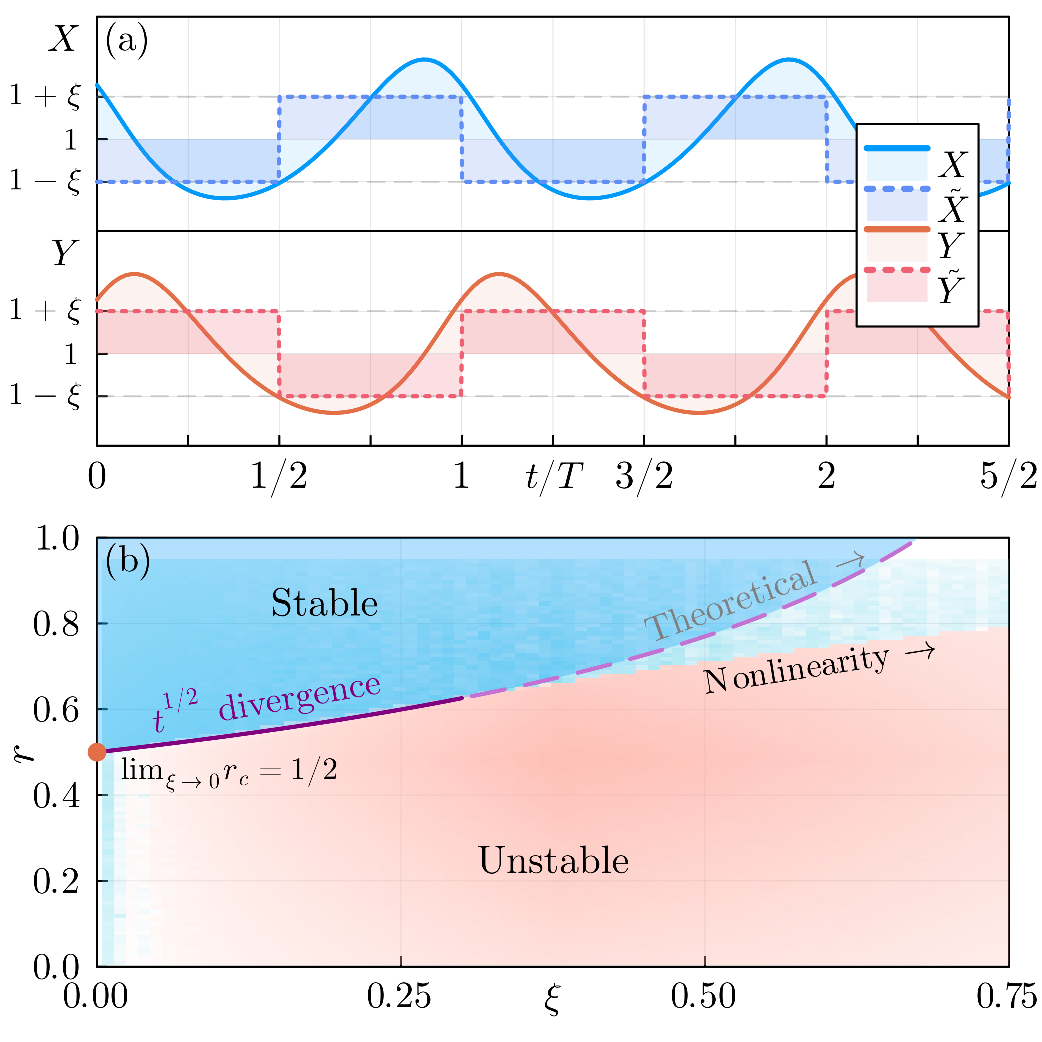}
\caption{(Color online) (a) Sketch of the step-function approximation where the periodic solution $[\bar{x}(t),\bar{y}(t)]^T$ are replaced by the step functions $[\tilde{x}(t),\tilde{y}(t)]^T$.    (b) the phase diagram obtained with the simplified model (separated by the dash line) and the numerical result of phase diagram (heatmap in the background). The heatmap displays $\sigma(t)$ after a long time ($t=600$), which remains as small as $\Delta$ (blue) for the stable phase and saturates to a large value of roughly 1 for the divergent phase (red).}
\label{fig:Fig3}
\end{figure}

\subsection{Step-function approximation}

The periodic solution [$\bar{x}(t),\bar{y}(t)]^T$ does not have a closed-form expression, thus it is impossible to analytically perform the time-ordering integral in Eq.(\ref{eq:Floquet0}) and derive an explicit form of the Floquet operator, even for a $2\times 2$ matrix. As we will show in the following, the qualitative dynamical behavior as well as the critical properties of our model do not crucially depend on the explicit formalism of the periodic function, what really matters is the amplitude and the period of the periodic function. Therefore, to analytically understand the different dynamical behavior and the transition between them, we adopt an approximation by replacing  the diagonal matrix in Eq.(\ref{eq:diag}) by a simplified formalism as (see Fig.\ref{fig:Fig3} a):
\begin{equation}
D(t)=
\begin{cases}
   \mathbb{I}+\xi \hat{\sigma}^z, \quad nT<t<(n+\frac 12)T\\
   \mathbb{I}-\xi \hat{\sigma}^z, \quad (n+\frac 12)T<t<(n+1)T
\end{cases},
\end{equation}
where $n$ is an integer, $\mathbb{I}$ represents a $2\times2$ identity matrix and $\hat{\sigma}^z$ denotes the z-component Pauli matrix. Furthermore, $\xi\in [0,1]$ characterizes the amplitude of the periodic oscillation, which is determined by the initial conditions in the original LV equationobtained by requiring that the step function share the same first order Fourier coefficient with the periodic solution $\bar{x}(t),\bar{y}(t)$:
\begin{equation}
  \int_0^T dt\,e^{-i\omega t} \cdot 2\xi\mathrm{sgn}(\sin{\omega t}) = \int_0^T dt\,e^{-i \omega t} [\bar{x}(t)-\bar{y}(t)],
\end{equation}
and if the nonlinearity is small so that harmonic approximation can by applied to $\bar{x}(t),\bar{y}(t)$, $\xi$ is simply promotional to the homogeneous oscillation amplitude:
\begin{equation}
  \xi = \frac{\pi}{8} \sqrt{ [ (\bar{x}(t)-x^\star)^2 + (\bar{y}(t)-y^\star)^2 ]}.
\end{equation}

\subsection{Quasi-energy band and the phase diagram of dynamical stability}

In the following, we demonstrate that despite the simplicity of such a step-function approximation, it can capture the essence of the non-Hermitian Floquet physics as well as the critical behavior, and explain the two different dynamics observed in the nonlinear Eq.(\ref{eq:LV}). By introducing $H_k^{\pm}=H_k^0 (\mathbb{I}\pm\xi \sigma_z)$, the evolution operator becomes
\begin{equation}
\mathcal{F}_k=e^{-i\frac T2 H_k^+}e^{-i\frac T2 H_k^-}=\begin{bmatrix}
      \frac{\cos\phi_k+\xi}{1+\xi}& -\frac{ie^{i\varphi_k}\sin\phi_k}{\sqrt{1-\xi^2}}      \\
     -\frac{ ie^{-i\varphi_k}\sin\phi_k}{\sqrt{1-\xi^2}}& \frac{\cos\phi_k-\xi}{1-\xi}
  \end{bmatrix},
\label{eq:Floquet1}
\end{equation}
where $\phi_k=\frac{\Delta_k T}2\sqrt{1-\xi^2}$ and $\Delta_k$ is the energy gap of $H_k^0$ ($\Delta_k=2\sqrt{(2+2\cos k)+2(1-\cos k)r^2 }$). $\varphi_k=\arg[-i(v+we^{-ik})]$. By diagonalizing the matrix presented in Eq.(\ref{eq:Floquet1}), one can obtain the eigenvalues of $\mathcal{F}_k$:
\begin{equation}
\lambda_k=\frac{\cos\phi_k-\xi^2\pm 2i|\sin\frac{\phi_k} 2|\sqrt{\cos^2\frac{\phi_k}2-\xi^2}}{1-\xi^2}.\label{eq:lambdak}
\end{equation}

Notably, the properties of $\lambda_k$  considerably  depend on the sign of $\cos^2\frac{\phi_k}2-\xi^2$, resulting in qualitatively different physical consequences.   If $\cos^2\frac{\phi_k}2>\xi^2$ for all the  k-modes, it is easy to check that $|\lambda_k|=1$, therefore we can introduce a real number $\theta_k\in [0,2\pi]$ such that $\lambda_k=e^{\pm i\theta_k}$.
Let $\varepsilon_k$ be the quasi-energy of the Floquet Hamiltonian  $H_k^F$, since $H_k^F=\frac iT \ln \mathcal{F}_k$, one can obtain $\varepsilon_k=\frac iT \ln \lambda_k=\mp \frac{\theta_k}T$. Therefore, in this case all the eigenvalues of the Floquet Hamiltonian $H_k^F$ are real and the dynamics of evolution remains stable. Consequently, there is no divergence for the deviation, and the dynamics is bounded within a finite regime around the homogeneous trajectory $[\bar{x}(t),\bar{y}(t)]$, agreeing with our numerical observation for large $r$. On contrast, when $\cos^2\frac{\phi_k}2<\xi^2$, $\lambda_k$ defined in Eq.(\ref{eq:lambdak}) becomes real and $|\lambda_k|\neq 1$. As a consequence, the eigenvalue of the Floquet Hamiltonian  $\varepsilon_k$ is no longer real, but with a pair of opposite imaginary parts, among which the one with positive imaginary part is responsible for the exponential divergence of the deviation observed in the case with small $r$. Obviously, such an exponential divergence predicted by the linear analysis cannot persist forever, because the nonlinear effect will finally take over and governs the long-time dynamics.

\begin{figure}[htb]
  \includegraphics[width=1.0\linewidth]{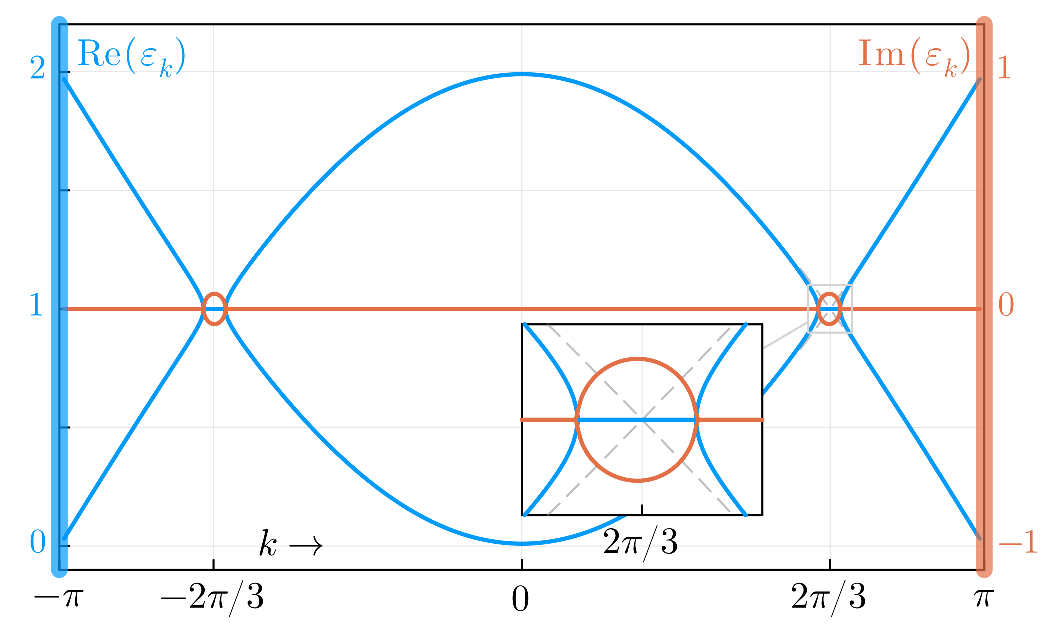}
  \caption{(Color online) Floquet quasi-energy band structure of a typical unstable case, where $\xi = 0.1, r = 0$. The inset magnifies the region where the real parts of quasi-energy become degenerate and the imaginary parts split into conjugate pairs.}
  \label{fig:Fig4}
\end{figure}

To illustratively address this mechanism, we numerically calculate one quasi-energy band in the unstable phase, see Fig.\ref*{fig:Fig4}. The imaginary parts of $\varepsilon_k$ are non-zero near $\Delta_k(k^*) = \omega$ which is just $k^* \approx 2\pi/3$. Any initial noise near $k^*$ gets amplified and exponentially grows. In contrast at relatively large $r$, if there is no such splitting of imaginary parts in the band, the dynamics stays quasi-unitary and stable.

It would also be interesting to analytically investigate the Lyapunov exponent of the divergence, named $\eta$, which corresponds to the maximum of imaginary part of the quasi-energy. Near $k^* = \mathrm{argmax}[\mathrm{Im}\,\varepsilon_k(k)]$, we introduce the detuning parameter $\nu = \frac{\omega}{\Delta_{k^*}}-1$ and neglect $\mathcal{O}(\xi^2)$ and smaller terms so that one can approximately obtain
\begin{equation}
  \varepsilon_k = \begin{cases}
    1 \pm \sqrt{\nu^2-\nu_c^2}, & |\nu| > \nu_c\\
    1 \pm i\sqrt{\nu_c^2-\nu}, & |\nu| \leqslant \nu_c
  \end{cases},
\end{equation}
where $\nu_c = 2\xi/\pi$ is proportional to $\xi$. To this first order approximation, $\eta = \nu_c = 2\xi/\pi$ and does not depend on $r$ (there is a tiny dependence on $r$ considering high order terms, and this approximation fails when $|r-r_c|$ is comparable with $\xi$ or the system is totally stable). This approximation agrees well with the inset of accurate calculation shown in Fig.\ref{fig:Fig4}.

Besides quantitatively explaining the Lyapunov component of divergence, we can further determine the critical condition for the system to be stable: The energy gap of $H_0^k$ satisfies $\Delta_k\in [4r,4]$ ($0<r<1$), which takes its minimum value $\Delta_{\min}=4r$ at $k=\pi$. Therefore, for $\xi$ fixed by small oscillation amplitude, the $\pi$-mode ($k=\pi$) will first become unstable as $r$ decreases below the critical value $r_c$ that satisfies $\cos[\pi r_c\sqrt{1-\xi^2}]=-\xi$, which indicates that $r_c\rightarrow \frac 12$ in the limit of $\xi\rightarrow 0$.

The phase diagram under this step-function approximation is also determined and plotted using smooth line in Fig.\ref{fig:Fig3} (b), where the phase boundary $r_c(\xi)$ is determined by the condition $\cos[\pi r_c\sqrt{1-\xi^2}]=-\xi$, at which the $\pi$-mode start to be unstable. The overlapped heatmap is the phase diagram from numerical simulation of the nonlinear GLVE Eq.(\ref{eq:LV}) and agrees with the approximation. Both results show that $r_c\rightarrow \frac 12$ when $\xi\rightarrow 0$, indicating that the approximation becomes exact in the limit of $\xi\rightarrow 0$ (but is still illustrative for any small $\xi$). For relatively large $\xi$, the nonlinearity cannot be neglected and leads to a shift of the boundary between two phase.

\begin{figure}[htb]
\includegraphics[width=0.99\linewidth]{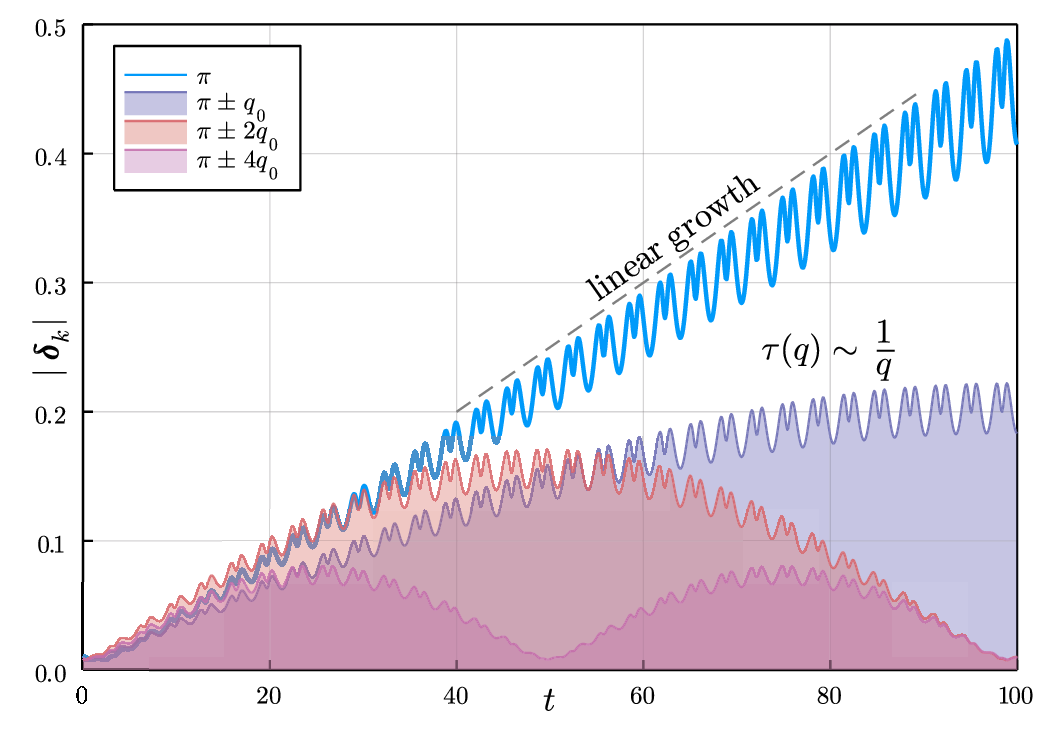}
\caption{(Color online)  The dynamics of $|\pmb{\delta}_k|$ for different k-modes that are right at or close to $k=\pi$. $q_0=\frac {\pi}{64}$}\label{fig:Fig5}
\end{figure}

\section{Critical dynamics: an emergent exceptional point}

In this section, we will explain the $t^{\frac 12}$ divergence of the average deviation $\sigma(t)$ observed right  at the critical point, which can be understood as a collective behavior of the k-modes close to $k=\pi$.
\begin{equation}
\sigma^2(t)=\frac 1L\sum_i \pmb\delta_i(t)\pmb\delta_i(t)=\frac 1L\sum_k \pmb\delta_k(t)\pmb\delta_{-k}(t) \label{eq:sigma}
\end{equation}
where the momentum summation is  over the k-mode in the first Brillonin Zone $k\in [0,2\pi]$ and  $\pmb\delta_k(t)=[\delta_k^x(t), \delta_k^y(t)]^T$.

\subsection{Dynamics of modes right at the exceptional point}

Right at the critical point, we first focus on the $\pi$-mode, whose dynamics at integer multiples of the period $T$ ($t=nT$) is governed by the Floquet operator
\begin{equation}
\mathcal{F}_\pi=2\xi\begin{bmatrix}
      1& -1      \\
      1& -1
  \end{bmatrix}+
  \begin{bmatrix}
      -1& 0      \\
     0& -1
  \end{bmatrix}.
  \label{eq:EP}
\end{equation}
Such a $2\times2$ matrix  has parallel eigenvectors with a degenerate eigenvalue $\lambda_\pi=-1$, indicating it is an exceptional point for the non-Hermitian matrix $\mathcal{F}_\pi$. Next, we will study the long-time dynamics governed by  $\mathcal{F}_\pi$.

The dynamics of $\pmb{\delta}_\pi(t)$ with $t=nT$ can be directly expressed as
\begin{equation}
\pmb{\delta}_\pi(nT)=\mathcal{F}^n_\pi \pmb{\delta}_\pi(0). \label{eq:dynamics}
\end{equation}
Assuming that initially $\pmb{\delta}_\pi(0)=[a,b]^T$, from Eq.(\ref{eq:dynamics}), one can derive that
\begin{equation}\label{eq:solutionpi1}
\pmb{\delta}_\pi(t)=(-1)^n \left\{ a\begin{bmatrix}
      1-Kt      \\
      -Kt
  \end{bmatrix}+
  b\begin{bmatrix}
      Kt      \\
     1+Kt
  \end{bmatrix}\right\},
\end{equation}
where $t=nT$, $K=\frac{2\xi}T$. In the long time limit $t\gg 1/K$, Eq.(\ref{eq:solutionpi1}) is reduced to:
\begin{equation}\label{eq:solutionpi2}
  \pmb{\delta}_\pi(t)= (b-a) Kt \begin{bmatrix}
      1      \\
      1
  \end{bmatrix},
\end{equation}
which indicates a linear divergence of  $|\pmb{\delta}_\pi(t)|$ at the critical point. This agrees very well with the numerical results as shown in Fig.\ref{fig:Fig5}, where the envelope of $|\pmb\delta_\pi(t)|$ growth linearly in time.

\subsection{Collective behaviour of modes and algebraic divergence}

For a single mode, the dynamics is either staying stable or diverging linearly, which indicates that the $1/2$ power law sublinear divergence is a collective behaviour under the thermodynamic limit. According to Eq.(\ref{eq:sigma}), all the k-modes contribute to $\sigma(t)$, while at the critical point, only the $\pi$-mode and those k-mode close to it dominate the long-time dynamics of  $\sigma(t)$. Now we focus on those k-modes close to $\pi$-mode with $k=\pi+q$ and $q\ll 1$. As shown in Fig.\ref{fig:Fig5}, for a k-mode that slight deviates from $k=\pi$, the envelope of $|\pmb\delta_{\pi+q}(t)|$ behavior resembles a sine function: initially, it grows linearly in time, while after a characteristic time scale $t^*_q$, it will significantly deviate from the linear function. Such a characteristic time scale is roughly a quarter of the period of the sine function, which in turn, is proportional to $1/|q|$, as shown in Fig.\ref{fig:Fig5}.

We can phenomenologically describe the dynamics of $\pmb{\delta}_{\pi+q}$ with
\begin{equation}
  |\pmb{\delta}_{\pi+q}(t)| = A_q \left|\frac{q_0}{q} \sin\left( \frac{q}{q_0}Kt \right) \right|,\label{eq:qgrowth}
\end{equation}
where $A_q$ is the an random amplitude but of the same order for all $q$. $q_0$ is a characteristic constant for all $q$. In this approximation, $t^*_q \sim Kq_0/|q| \sim |q|^{-1}$. Also, the linear growth of $\pi$-mode is recovered in the limit that $q\rightarrow 0$.

Qualitatively, the closer a k-mode is to $k=\pi$, the longer it can contribute a linear component to $\sigma(t)$. At a fixed time $t$, only $\mathcal{N}(t)\sim 1/t$ of those k-modes satisfy $t^*_q>t$ and are still linearly growing, which explains why the collective dynamics of $\sigma(t)$ is sublinear. Quantitatively, by substitute the phenomenological expression for $\pmb{\delta}_{\pi+q}(t)$ Eq.(\ref{eq:qgrowth}), we can explicitly calculate $\sigma(t)$:
\begin{equation}
\begin{split}
  \sigma^2(t) & = \int_{-\pi}^{+\pi} dq\,\rho(q) |\pmb{\delta}_{\pi+q}(t)|^2\\
   \approx\langle A_q^2 & \rangle \frac{N}{2\pi} Ktq_0 \int_{-\infty}^{+\infty} dx \, \frac{\sin^2{x}}{x^2} = \frac{NKq_0}{2} \langle A_q^2 \rangle \cdot t
\end{split},
\end{equation}
where the amplitude $A_q$ is assumed to be uniform over all $q$ and replaced by its average $\langle A_q \rangle$ over $q$. Therefore, one can obtain $\sigma(t) \sim t^{1/2}$, which agree with the critical power law divergence of the nonlinear GLVE.


\section{Conclusion and outlook}

In summary, this study show that non-Hermitian physics, which used to be considered as a consequence of dissipative quantum systems,  can emerge in classical non-linear systems out of equilibrium. This work also provide a new member to the quasi-Hermitian family with real eigenvalues.   It is shown that the interplay between temporal periodicity and non-Hermicity can lead to intriguing dynamic behaviors\cite{Li2019,Longhi2017,Koutserimpas2018,Zhou2018,Zhou2019,Hockendorf2019,Wu2020,Zhang2020}.

We also point out that the expansion technique Eq.(\ref{eq:expansion}) can be applied to other predator-prey type GLVE and results a Hamiltonian like Eq.(\ref{eq:Hk}) that is usually time-dependent and non-Hermitian, see Appendix A. Our method also provides an opportunity to understand the phenomena such as the pattern formation\cite{Menezes2021} and phase coexistence\cite{Knebel2013} in GLVE from a perspective of non-Hermitian physics. .


\section*{Appendix A: derivation of time-dependent non-Hermitian Hamiltonians from generic GLVE}

Mathematically, GLVE can be written in the generic form where all variables and parameters are real-valued:
\begin{equation}\label{eq:genericGLVE}
  \dot{x}_i = x_i \left(\gamma_i + \sum_{j\ne i} \kappa_{ij} x_j \right),
\end{equation}
where $x_i$ denotes the mass on site $i$ and is usually considered positive. $\gamma_i$ is the corresponding growth/decay rate. The coupling coefficients $\kappa_{ij}$ characterize the nonlinear interaction among sites.

Now we focus on the evolution of perturbation $\delta_i(t)$ on a given solution $X_i(t)$ (not necessarily periodic or stationary). Substitute $x_i(t) = [1+\delta_i(t)]X_i(t)$, we get
\begin{equation}
  (1+\delta_i)\dot{X}_i + X_i \dot{\delta}_i = (1+\delta_i)X_i \left[\gamma_i + \sum_{i\ne j}\kappa_{ij} X_j(1+\delta_j)  \right],
\end{equation}
and by neglecting $o(\delta^2)$ terms like $\delta_i\delta_j$, we obtain a EOM for $\delta_i$ that does not explicitly contain $\gamma_i$:
\begin{equation}\label{eq:appendix1}
  \dot{\delta}_i = \kappa_{ij} X_j \delta_j.
\end{equation}

Now let's use the following more heuristic symbols
\begin{equation}
  D_{ij}(t) = X_i(t) \delta_{i,j}, \quad \{H_0\}_{ij} = i \kappa_{ij},
\end{equation}
where $D = \mathrm{diag}[X_1(t)\cdots X_n(t)]$ is a diagonal matrix. Now we multiply EOM Eq.(\ref{eq:appendix1}) by a factor of $i$. Then it turns out to be
\begin{equation}\label{eq:appendix2}
  i \frac{d\delta_i}{dt}  = \{H_0\}_{ij} D_{jk} \delta_k,
\end{equation}
or
\begin{equation}\label{eq:appendix3}
  i \frac{d}{dt} \pmb{\delta} = H_0 D(t) \pmb{\delta}.
\end{equation}
which is essentially a single-particle Schrodinger equation with a time-dependent non-Hermitian "Hamiltonian"
\begin{equation}
  H(t) = H_0 D(t)
\end{equation}

For predator-prey models, $\kappa_{ij}$ are sign-constrained that $\kappa_{ij}\kappa_{ji}<0$ and is called antagonistic\cite{Mambuca2022}, where the antisymmetric ($\kappa_{ij}=-\kappa_{ji}$) case is often of interest \cite{Knebel2013,Knebel2020,Umer2022}. If the latter is true, then $H_0^\dagger = H_0$ and $H_0$ will be Hermitian. Moreover, the generic GLVE Eq.(\ref{eq:genericGLVE}) can be written as
\begin{equation}\label{eq:genericGLVE2}
  \dot{y}_i = \gamma_i + \sum_{j\ne i} \kappa_{ij} \exp{y_j},
\end{equation}
where $y_i = \log x_i$, where we can infer that $\forall X_i(t)$ will stay positive as long as $\forall X_i(t=0)>0$. Therefore, $D(t)$ is positive semidefinite and Cholesky factorization $L^\dagger L = D$ is well-defined with $L = \sqrt{D}$. It is easy to check that $H=H_0 D$ is similar to another Hermitian Hamiltonian $\mathcal{H} = L^\dagger H_0 L$:
\begin{equation}
  H =  (L^\dagger)^{-1} \mathcal{H} L^\dagger.
\end{equation}

This guarantees that $H$ share the same eigenvalues $\{\omega_i\}$ with $\mathcal{H}$, which are real; their eigenvectors ($\{\psi_i\}$ for $H$ and $\{\phi_i\}$ for $\mathcal{H}$) are usually different, but can be related by the transformation:
\begin{equation}
  \phi_i = L^\dagger \psi_i = \sqrt{D} \psi_i.
\end{equation}

Since $\det \sqrt{D} = \sqrt{\prod_{i=1}^N X_i(t)} > 0$, the inverse transformation
\begin{equation}
  \psi_i = (L^\dagger)^{-1} \phi_i = D^{-\frac 1 2} \phi_i
\end{equation}
is well-defined and keeps the span $\{\psi_i\}$ non-degenerate.

If one perform such expansion around a saturate solution $X_i(t) = X^\star_i$, then $H$ is time-independent. Despite the non-Hermicity of $H$, this will not lead to more intriguing dynamics than $H'$. One would expect quasi-unitary dynamics and will not encounter exceptional points because the non-degeneracy of $\{\psi_i\}$ means that none of the eigenvectors is parallel to another.

On contrast, non-trivial dynamics lies behind the time-dependence of $H(t)$. If $[H(t_1),H(t_2)]\ne 0$, then the effective Hamiltonian on a given time interval can possibly be PT-broken with complex eigenvalues or hosts exceptional points with parallel eigenvectors, exhibiting non-trivial dynamics. Additionally, Floquet analysis can be applied if the solution $X_i(t)$ is periodic.

\section*{Acknowledgments}

This work is supported by the National Key Research and Development Program of China (Grant No. 2020YFA0309000), NSFC of  China (Grant No.12174251), Natural Science Foundation of Shanghai (Grant No.22ZR142830),  Shanghai Municipal Science and Technology Major Project (Grant No.2019SHZDZX01). ZC thank the sponsorship from Yangyang Development Fund.


%

\end{document}